# CARLA: Adjusted common average referencing for cortico-cortical evoked potential data

Harvey Huang*[a], Gabriela Ojeda Valencia[b], Nicholas M. Gregg[c], Gamaleldin M. Osman[c,f], Morgan N. Montoya[b], Gregory A. Worrell[b,c], Kai J. Miller[b,d], Dora Hermes*[b,c,e]

[a]Mayo Clinic Medical Scientist Training Program, Rochester, MN; [b]Department of Physiology and Biomedical Engineering, Mayo Clinic, Rochester, MN; [c]Department of Neurology, Mayo Clinic, Rochester, MN; [d]Department of Neurologic Surgery, Mayo Clinic, Rochester, MN; [e]Department of Radiology, Mayo Clinic, Rochester, MN, 55901. [f]Division of Child Neurology, Department of Pediatrics, McGovern Medical School at UTHealth, Houston, TX.

*Corresponding: huang.harvey@mayo.edu & hermes.dora@mayo.edu

Declaration of Competing Interest: Unrelated to this research, G. A. Worrell has licensed intellectual property developed at Mayo Clinic to Cadence Neuroscience Inc. and NeuroOne Inc. All other authors declare that they have no known competing financial interests or personal relationships that could have appeared to influence the work reported in this paper.

## CRediT authorship contribution statement

**Harvey Huang:** Conceptualization, Data curation, Formal Analysis, Funding Acquisition, Investigation, Methodology, Software, Visualization, Writing - Original Draft, Writing - Review & Editing. **Gabriela Ojeda Valencia:** Data curation, Validation. **Nicholas M. Gregg:** Investigation, Writing - Review & Editing. **Gamaleldin M. Osman:** Investigation, Writing - Review & Editing. **Morgan N. Montoya:** Data curation, Writing - Review & Editing. **Gregory A. Worrell:** Funding Acquisition, Project Administration, Resources. **Kai J. Miller:** Methodology, Resources. **Dora Hermes:** Conceptualization, Funding Acquisition, Investigation, Methodology, Project Administration, Resources, Writing - Original Draft, Writing - Review & Editing.

# Abstract

Human brain connectivity can be mapped by single pulse electrical stimulation during intracranial EEG measurements. The raw cortico-cortical evoked potentials (CCEP) are often contaminated by noise. Common average referencing (CAR) removes common noise and preserves response shapes but can introduce bias from responsive channels. We address this issue with an adjusted, adaptive CAR algorithm termed "CAR by Least Anticorrelation (CARLA)".

　　CARLA was tested on simulated CCEP data and real CCEP data collected from four human participants. In CARLA, the channels are ordered by increasing mean cross-trial covariance, and iteratively added to the common average until anticorrelation between any single channel and all re-referenced channels reaches a minimum, as a measure of shared noise.

　　We simulated CCEP data with true responses in 0 to 45 of 50 total channels. We quantified CARLA's error and found that it erroneously included 0 (median) truly responsive channels in the common average with ≤42 responsive channels, and erroneously excluded ≤2.5 (median) unresponsive channels at all responsiveness levels. On real CCEP data, signal quality was quantified with the mean $R^2$ between all pairs of channels, which represents inter-channel dependency and is low for well-referenced data. CARLA re-referencing produced significantly lower mean $R^2$ than standard CAR, CAR using a fixed bottom quartile of channels by covariance, and no re-referencing.

　　CARLA minimizes bias in re-referenced CCEP data by adaptively selecting the optimal subset of non-responsive channels. It showed high specificity and sensitivity on simulated CCEP data and lowered inter-channel dependency compared to CAR on real CCEP data.

# Keywords

# 1. Introduction

## 1.1. Offline re-referencing of intracranial EEG data

Intracranial brain stimulation is used increasingly to map effective connectivity in the human brain (Friston, 1994; Keller et al., 2014). A fundamental understanding of how electrical stimulation spreads through brain networks can be acquired by applying single pulse stimulation to select pairs of intracranial EEG (iEEG) electrodes and measuring the evoked potentials at others (Araki et al., 2015; Keller et al., 2014; Kundu et al., 2020). The measurement electrodes may lie on the convexity of the brain surface from electrocorticography (ECoG) or in deeper structures from stereoelectroencephalography (stereoEEG, sEEG). Single pulse electrical stimulation is performed using subsets of these electrodes, most often in anode-cathode pairs, in cortical and non-cortical structures such as white matter and thalamus. The voltage peaks and troughs captured by the measurement electrodes, termed cortico-cortical evoked potentials (CCEPs), quantify synchronous signal propagation downstream from the stimulation site and are a macroscopic analog to local field potentials (Logothetis et al., 2010; Silverstein et al., 2020). Accurately characterizing the amplitudes and shapes of these CCEPs can provide valuable information on connectivity between stimulation and measurement areas (Barbosa et al., 2022; Huang et al., 2023; Krieg, 2017; Kundu et al., 2020; Matsumoto et al., 2004, 2007; Valencia et al., 2022).

Raw iEEG data at each channel are recorded as the amplified potential difference between each measurement electrode and a reference electrode, which is chosen *a priori* by the clinician or researcher. The reference electrode is either external, in which case it is located on the skin or scalp, or internal, in which case it is another iEEG electrode that ideally picks up as little neural activity as possible. In both situations, the amplified iEEG data is prone to contain broadband and periodic noise (e.g., 50 or 60 Hz "line noise") (Mercier et al., 2022; Nunez & Srinivasan, 2006). The amplitude of this noise can be several orders of magnitude greater than the physiological signal of interest. Furthermore, the time-varying potential differences picked up by the reference electrode are synchronously introduced to all channels. In general, these factors mean that the signal quality of iEEG data can be substantially improved by offline re-referencing prior to interpretation.

There is no single perfect re-referencing method for iEEG data, as different signal features are better preserved by different references. Local re-referencing methods such as bipolar and Laplacian, which are calculated as the differences between spatially adjacent sets of electrodes, attenuate signal features that are shared between neighboring electrodes to magnify highly focal signal features such as broadband activity (Dickey et al., 2022; Li et al., 2018; Mercier et al., 2022; Shirhatti et al., 2016). In the case of CCEP data, large amplitude deflections can be distributed across multiple neighboring electrodes and are often of greatest interest to quantify. Local re-referencing methods are therefore inadequate in this context because they can markedly reduce the amplitude of CCEPs, distort temporal profiles, and introduce phase reversals (Arnulfo et al., 2015; Shirhatti et al., 2016). A less severe re-referencing method for CCEP data is the common average reference (CAR), in which the mean signal across all channels approximates the common noise and is subtracted from each channel. One major disadvantage to this approach is that signals from highly responsive channels may be smeared into all other channels and introduce bias. Given these potential issues with both local re-referencing methods and CAR, many CCEP studies opt out of re-referencing altogether. Recently, a few modified referencing methods based on CAR have been proposed. One proposal is to use only channels corresponding to white matter electrodes, with the assumption that responses are negligible at white matter electrodes (Arnulfo et al., 2015; Uher et al., 2020). This assumption, however, does not always hold true (Mercier et al., 2017; Uher et al., 2020), and the number of white matter electrodes available can vary depending on the clinical plan. Using anatomical data to inform iEEG data preprocessing also requires accurate coregistration and segmentation of imaging data. Another proposal is to only use channels with low-variance signals as a reference. This approach requires determining which channels qualify as "low-variance", which can differ by dataset and by stimulation site within the same dataset. In this work, we center on the concept of low variance channels to develop a data-driven re-referencing algorithm, which adaptively optimizes which channels to include in the common average based on the input CCEP data. We demonstrate that our algorithm effectively separates responsive from non-responsive channels on simulated CCEP data and performs well on real CCEP data recorded from four human subjects, as quantified by a reduction in cross-channel explained variance.

## 1.2 Theoretical components of cortico-cortical evoked potential data

The goal of common average referencing is to improve the signal of the evoked potential relative to the noise, without introducing bias. We first use simulated CCEP data to introduce the theoretical CCEP components of interest (Figure 1). The raw time series present at a single channel is the sum of a transient stimulation artifact, an evoked potential (when present), a random noise component, and a common noise component. The evoked potential is a set of consistent stimulation-locked voltage deflections across trials. The random noise component consists of brown noise that is unique to each channel and trial. The common noise is shared across all channels but unique in time to each trial, which we model as a sum of periodic line noise (60Hz + harmonics) and Brown noise, which follows a $1/f^2$ power spectrum (Miller et al., 2009).

      The optimal common average referencing procedure should remove the common noise but minimally introduce bias from responsive channels. The random noise is not removed by re-referencing; rather, it is mitigated by averaging the signal across multiple trials.

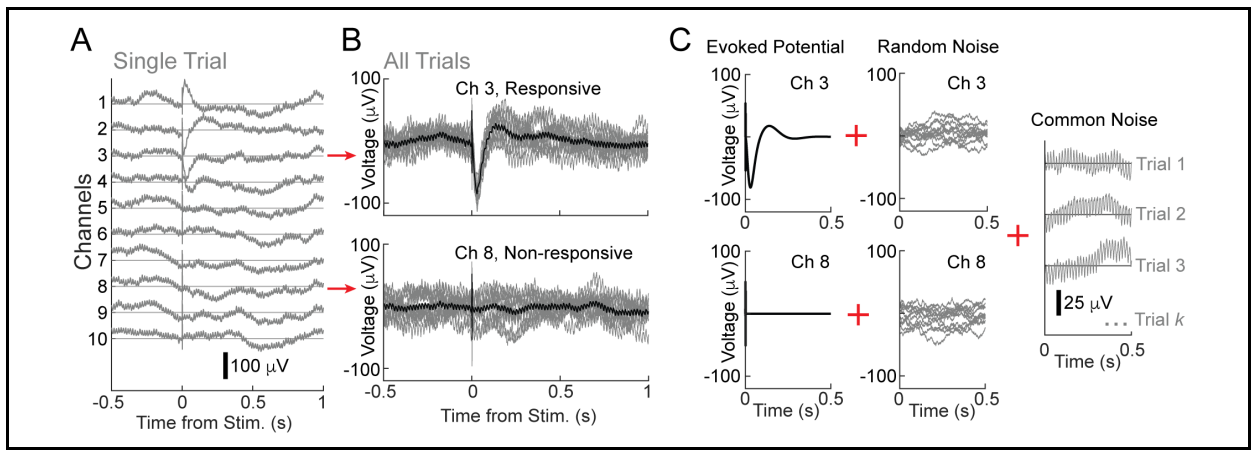

> **Figure 1. Theoretical components of the CCEP data. A,** Simulated CCEP data for 10 channels at one trial. Channels 1 through 4 contain true, distinct evoked potential responses, while channels 5 through 10 are non-responsive. **B,** 12 individual trials shown for channels 3 and 8 in gray, with the mean across all trials in black. **C,** The three components of the CCEP data shown for channels 3 and 8. This includes a responsive evoked potential in channel 3 but not 8, random noise (brown noise) unique to each channel, and common noise (60 Hz + harmonics + brown noise) shared across channels. The stimulation artifact is between 0 and 2 ms and depicted here at the onset of the evoked potential.

## 2. Materials and Methods

### 2.1. Ethics statement

The experiments in this study were conducted according to the guidelines of the Declaration of Helsinki and approved by the Institutional Review Board of the Mayo Clinic (IRB #15-006530), which also authorizes sharing of the data. Each patient/representative voluntarily provided independent written informed consent/assent to participate in this study.

### 2.2. Analytic framework

The CAR is prone to introducing bias from highly responsive channels into all other channels. Variations on CAR exclude responsive channels based on anatomical position or signal properties. We propose here a variation on CAR that involves calculating the signal variance on a predetermined time interval (i.e., between 10 and 300 ms) for all channels, and then including only a subset of channels with the lowest variance in the common average. Signal variance is equivalent to the energy of the mean-centered signal on the time interval of interest, and is therefore related to response strength. To determine how many channels should be included in the (adjusted) common average, a percentile threshold can be chosen arbitrarily or by visual inspection of plotted signals, as an upper bound on variance. If the threshold is too lenient (e.g., 95% of all channels by lowest variance included in the common average), the common average risks including bias from responsive channels when the stimulation site projects to many channels. If the threshold is too strict (e.g., 10% of all channels by lowest variance included), there may be non-negligible dependencies between the common average and the random noise of each channel used to create it (Box 1).

> **Box 1. Theoretical basis to optimize the number of channels in a common average.** Let $S$ be the true common noise across all channels. $S$ can be thought of in this context as a common "signal" of interest, and the goal of CAR is to best estimate $S$ from a sample of channels and subtract it from all channels. $S$ is estimated by averaging over $n$ channels. Even if the $n$ channels are not responsive, they each contribute $S$ plus channel-specific random noise to the common average. Assuming random noise with 0 mean that is statistically independent across the $n$ channels, the signal-to-noise ratio of the common average is proportional to $\sqrt{n}$ (Drongelen, 2018). Therefore, $S$ is best estimated by increasing the number of channels used to construct the common average, up to the point when evoked signals of interest from responsive channels would be introduced into the common average as a systematic bias. Thus, the problem of determining the correct number of channels to use for CAR is suited for optimization.

Our algorithm optimizes the number of channels to include in the adjusted common average, adaptively based on the signal features in the input data set. We considered the following set of criteria in developing this algorithm:
1. Logically similar to the rationale used in visual inspection of signals
2. Depends only on the CCEP data, not on other information such as anatomical location
3. Effective across a wide range of cases containing sparse to widespread channel responses
4. Minimal dependency on statistical thresholds and user-input parameters

## 2.3. Algorithmic optimization of channels in common average

The proposed algorithm, which we refer to as CAR by Least Anticorrelation (CARLA), is described below and a single-trial example is shown in Figure 2. In brief, the goal is to include the *n* channels with the lowest signal variance or covariance to construct the common average, on a per-trial basis, such that there is the least amount of anticorrelation between the re-referenced channels and any single channel that makes up the common average. CARLA is best applied after data have been preprocessed to remove channels and trials containing large artifacts (see section 2.5).

1. *Input signals*: Denote the input to be $V_{input}$, a 3-dimensional matrix of *N* channels by *T* time points by *K* trials conducted for the same experimental condition (e.g., same stimulation site).
2. *Essential preprocessing*: We filter $V_{input}$ with notch filters at 60 Hz and its first two harmonics (120 Hz, 180 Hz), and denote the filtered signals *V* (Figure 2A). Filtering is done such that signal variance (energy) and cross-channel correlations in *V* are more directly reflective of the physiologic responses and less influenced by line noise, which may be of high amplitude.
3. *Ordering channels to include in the common average*: All channels are ranked in order of increasing mean covariance across trials (self-covariance excluded) in *V* on a responsive time interval. Covariance is a good estimator of both signal strength and reliability across trials, with an expected value of 0 when trials are uncorrelated. Variance is used in place of covariance in the single-trial situation ($K$ = 1).
    a. *CCEP-specific parameter*: We chose the response interval to be 10 ms to 300 ms post-stimulation, as this window reasonably reflects early, direct responses (e.g., N1) that are typically contained in the first 100 ms, as well as later, indirect responses that typically begin well before 300 ms (e.g., N2). The first 10 ms are omitted in order to avoid the stimulation artifact. This interval may be adjusted based on prior estimates of response duration.
4. *Anticorrelation as the statistic of interest*: Some degree of anticorrelation is expected to exist between the common average referenced channels and each original channel (prior to re-referencing) that makes up the common average, because re-referencing distributes the random noise at each channel into all other channels. As the number of non-responsive channels used to construct the common average grows, the anticorrelation between any individual channel and all other re-referenced channels decreases in magnitude, because the proportion of each individual channel in the common average decreases $\propto$ *1/n*. However, when a channel with a large response that deviates strongly from the common noise, such as a CCEP, is incorporated into the common average, it introduces a bias into the re-referenced channels that manifests as a strong anticorrelation between itself and all other re-referenced channels without a similar response morphology. CARLA optimizes the number of channels to include in the common average by minimizing this anticorrelation.

5. *Iterative calculation of the statistic*: For *n* = 2, 3, … *N*, the *n* channels with the lowest (co)variance are iteratively taken to form subset $U_n$, $U_n \subseteq V$. At each iteration, the common average is calculated from $U_n$, separately for each trial, and subtracted from $U_n$ to yield the re-referenced $U_{n,CAR}$ (Figure 2B). Then for *i* = 1 … *n*, $\bar{z}_{i,n}$ denotes the average Fisher z-transformed Pearson's r between the *i*th channel in $U_n$ (before re-referencing) and all other channels in $U_{n,CAR}$ on the response interval. $\bar{z}_{i,n}$ quantifies the average similarity of the response at the *i*th channel to the responses at all other channels in the subset when re-referenced. At each *n*, the most negative $\bar{z}_{i,n}$ over all *i* = 1, 2, … *n*, min($\bar{z}_{i,n}$), belongs to the most globally anticorrelated channel in $U_n$, and is designated $\zeta$. When *K* > 1, $\zeta$ is estimated for the mean across trials by bootstrapping.

6. *Optimization criteria*: The optimal common average can be constructed from the *n* channels with the lowest (co)variance, at *n* corresponding to the least negative value (global maximum) in $\zeta$ (Figure 2C). This occurs when there is the least average anticorrelation between any single channel in $U_n$ and all others in $U_{n,CAR}$. Alternatively, a more sensitive *n* may be identified at the first local maximum in $\zeta$ prior to a significant decrease, where significance is calculated by bootstrapping across trials (when *K* > 1). This stopping criterion is explained in greater detail in the results section below, and it better identifies the smallest subset of channels that does not contain any responsive channels.

7. *Output signals*: The final common average is calculated per trial as the mean time series across the optimal *n* channels with the lowest (co)variance from the unfiltered $V_{input}$. The common average is subtracted from $V_{input}$ to produce the re-referenced signals.

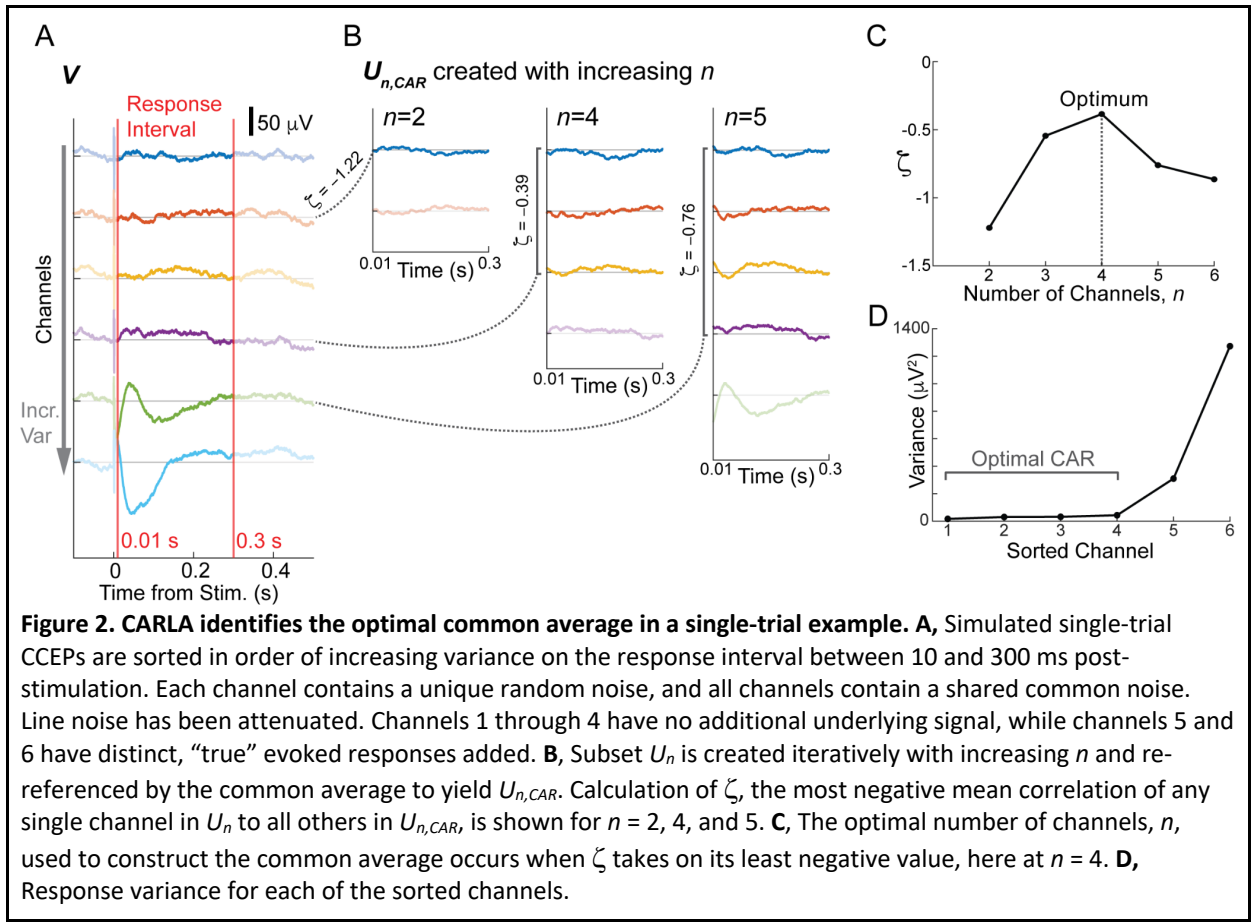

**Figure 2. CARLA identifies the optimal common average in a single-trial example. A,** Simulated single-trial CCEPs are sorted in order of increasing variance on the response interval between 10 and 300 ms post-stimulation. Each channel contains a unique random noise, and all channels contain a shared common noise. Line noise has been attenuated. Channels 1 through 4 have no additional underlying signal, while channels 5 and 6 have distinct, "true" evoked responses added. **B,** Subset $U_n$ is created iteratively with increasing $n$ and re-referenced by the common average to yield $U_{n,CAR}$. Calculation of $\zeta$, the most negative mean correlation of any single channel in $U_n$ to all others in $U_{n,CAR}$, is shown for $n$ = 2, 4, and 5. **C,** The optimal number of channels, $n$, used to construct the common average occurs when $\zeta$ takes on its least negative value, here at $n$ = 4. **D,** Response variance for each of the sorted channels.

Figure 2B shows that when too few channels are used to construct the common average (e.g., $n$=2), re-referenced signals show strong anticorrelation with any individual channel making up the common average. When a channel with high variance is incorporated into the common average (e.g., $n$=5), re-referenced signals show strong anticorrelation with the high-variance channel due to introduced bias. The optimal common average, at $n$=4 in this case, includes all non-responsive channels and excludes highly responsive channels. Intuitively, this cutoff co-occurs with a shoulder in the sorted channel variances (Figure 2D).

Note also from Figure 2B that $\zeta$ at each $n$ often belongs to the $n$th and most recently added channel ($i = n$) in $U_n$. Since the $n$th channel has the highest variance out of all channels in $U_n$ (or covariance across trials when $K > 1$), it is most likely to have the greatest influence on the common average. However, if the $n$th channel is anticorrelated with earlier responsive channels, it may paradoxically exhibit a positive correlation with the re-referenced channels on average, resulting in spuriously positive $\bar{z}_{i,n}$, $i=n$. This motivates the necessity of finding $\min(\bar{z}_{i,n})$ out of all possible channels at each iteration of $U_n$ to designate as $\zeta$.

## 2.4. Construction of simulated CCEP data

CARLA was first tested on simulated CCEP data, where we varied the number of channels containing true responses. Simulated CCEP responses were created at 4800 Hz for each channel and trial by summing the following components (see Figure 1): an evoked potential, a random noise unique to each trial and channel, a common noise shared across channels but unique to each trial, and a transient stimulation artifact at the beginning of the signal. Non-responsive channels were created in the same way but with evoked potential equal to 0. The simple simulation procedure is described sequentially below. Note that our goal here was to generate the necessary components to test CARLA, rather than to provide the best biophysical approximation of real CCEPs.

Each evoked potential, $P(t)$, was modeled as a sum of two causal sinusoids, each enveloped by a difference of two exponential decays (Figure 3):

$$P(t) = A \times [S_1(t) + S_2(t)], t \geq 0$$
$$\text{where } S_1(t) = (e^{-t/\tau_1} - e^{-t/\tau_2}) \times \sin(2\pi f_1 t - \phi_1)$$
$$\text{and } S_2(t) = (e^{-t/\tau_3} - e^{-t/\tau_4}) \times \sin(2\pi f_2 t - \phi_2).$$

$t$ is the time from stimulation, in seconds. $A$ is a scalar amplitude, sampled uniformly at random between 80 and 120. $\tau_1$ and $\tau_3$ are time constants that control how quickly each envelope decays, sampled uniformly at random between 0.01 and 0.03 and between 0.06 and 0.14 respectively. $\tau_2$ and $\tau_4$ determine the rise time of each envelope and were held constant at 0.005 and 0.025, respectively. $f_1$ and $f_2$ are the frequencies of each sinusoid in Hz, sampled uniformly at random between 8 and 12 and between 1 and 3, respectively. Finally, $\phi_1$ and $\phi_2$ are the phases of each sinusoid, each sampled uniformly at random between 0 and $2\pi$ (all possible phases). The intervals used for the time constants and frequencies were determined empirically from standard CCEP characteristics. $S_1$ captures a fast, direct response, with a mean half-period of 50 ms similar to the CCEP N1 deflection. $S_2$, the indirect response, has a half-period and exponential time constants that are 5 times slower than $S_1$, on average, each of which also includes more jitter. The resulting evoked potentials are diverse in waveform, characteristic of those recorded at different layers of cortex by sEEG (Huang et al., 2023; Miller et al., 2021).

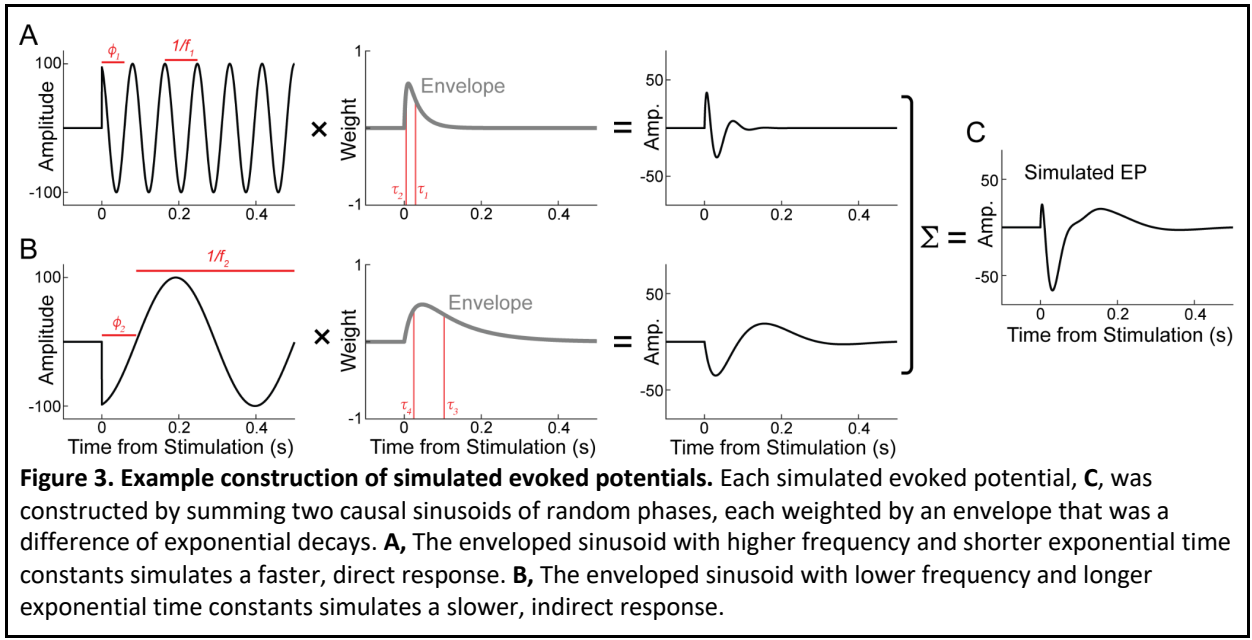

**Figure 3. Example construction of simulated evoked potentials.** Each simulated evoked potential, **C**, was constructed by summing two causal sinusoids of random phases, each weighted by an envelope that was a difference of exponential decays. **A,** The enveloped sinusoid with higher frequency and shorter exponential time constants simulates a faster, direct response. **B,** The enveloped sinusoid with lower frequency and longer exponential time constants simulates a slower, indirect response.

The random noise unique to each channel and trial was brown noise produced by cumulatively summing independent samples from a standard normal distribution, then multiplied by a gain factor of 0.4. This was then high pass-filtered with a Butterworth filter above a -3 dB cutoff frequency of 0.5 Hz by forward-reverse filtering. This filtering step is the same as applied to our real CCEP data to remove low-frequency drift.

The common noise for each trial consisted of two subcomponents summed together. The first subcomponent is a periodic line noise component, $L(t)$, that is the sum of a 60 Hz sinusoid plus two out-of-phase harmonics at 120 Hz and 180 Hz:

$$L(t) = A_1 S_1(t) + A_2 S_2(t) + A_3 S_3(t),$$
where $S_1(t) = sin(2\pi 60 t - \phi_1),$
$S_2(t) = sin(2\pi 120 t - \phi_2),$
and $S_3(t) = sin(2\pi 180 t - \phi_3).$

$t$ is time from stimulation, in seconds. $A_1$, $A_2$, and $A_3$ are scalar amplitudes equal to 8, 2, and 1, respectively, representing decreasing contributions from the two harmonics. $\phi_1$, $\phi_2$, and $\phi_3$ are phases each sampled uniformly at random between 0 and $2\pi$. The second subcomponent is brown noise constructed the same way as the random noise above.

Finally, a sharp transient stimulation artifact, $Q(t)$, was also added between 0 and 2 ms post-stimulation, as a fast constant-phase sinusoid described by:

$$Q(t) = A \times sin(2\pi 600 t), 0 \leq t < 0.002.$$

$t$ is time from stimulation, in seconds, and $A$ is a scalar amplitude sampled uniformly at random between 47 and 53. This artifact was included for completeness but did not affect the results since it was well before the analyzed response window of [10, 300 ms].

Simulated CCEP data with 50 channels and 12 trials were created at 46 levels of responsiveness, where 0 (0%) to 45 (90%) of all channels contained true responses. 30 random sets of CCEPs were simulated at each level of responsiveness. The performance of CARLA was evaluated on these simulated data by quantifying the error in CARLA's output. For each simulated set, a false negative (FN) indicates a responsive channel erroneously included in the common average and a false positive (FP) indicates a non-responsive channel erroneously excluded from the common average.

## 2.5. Collection and preprocessing of real CCEP data

iEEG voltage data were measured from 4 human subjects (2 male, 2 female, see Table 1) who had been implanted with stereo EEG (sEEG) electrodes for epilepsy monitoring. Recorded data were digitized at 4800 Hz on a G.Tec G.HIAmp biosignal amplifier, and then high pass-filtered above 0.5 Hz by forward-reverse filtering with a second-order Butterworth filter (MATLAB filtfilt).

Electrode pairs were stimulated 12 times (trials) each with a single biphasic pulse of 200 μs pulse width and 6 mA amplitude every 3-5 s, using the G.Tec G.Estim PRO electrical stimulator. Stimulation sites were excluded from analysis if they overlapped seizure onset zones, per physician records. Measurement electrodes were excluded from analysis if they were located outside of the brain or if they contained consistent artifacts throughout an experimental run. Table 1 lists the number of measurement electrodes and the number of stimulation sites after the above criteria have been applied. Then for each individual stimulation site analyzed, measurement electrodes were excluded for those trials if they were part of the stimulation pair or located on the same lead and within 2 electrodes of the stimulated pair (inter-electrode distance was 3.5 mm, on average). Finally, all trials were visually inspected; for each individual stimulation site, measurement electrodes were excluded when they contained artifact or excessive epileptiform activity in any trial, and single trials were excluded altogether when the artifact or epileptiform activity was present across many measurement electrodes. This resulted in a minimum of 10 trials included in subsequent analysis for two stimulation sites in subject 3, and 11 or 12 trials included for all other stimulation sites.

**Table 1. Subject Demographics**

| Subject ID | Age | Sex | Hemisphere(s) implanted | No. of measurement electrodes | No. of stimulation sites |
| --- | --- | --- | --- | --- | --- |
| 1 | 18 | M | Bilateral | 213 | 32 |
| 2 | 19 | M | Bilateral | 204 | 16 |
| 3 | 19 | F | Left | 190 | 23 |
| 4 | 16 | F | Right | 114 | 11 |

A list of age, sex, implanted hemisphere(s), number of measurement sEEG electrodes, and number of stimulation sites for each subject.

## 2.6. Quality metric of referencing on real CCEP data

In contrast to simulated CCEP data, the true responsiveness of channels in real CCEP data is unknown. Therefore, we quantified the quality of re-referenced signals using the mean cross-channel coefficient of determination ($R^2$), and we used this metric to compare referencing performance across different versions of CAR. In well-referenced CCEP data, the average cross-channel $R^2$ should be relatively low as there are minimal synchronous interdependencies between non-responsive channels in the data. In poorly referenced data, the average $R^2$ is inflated by positive correlations due to the common noise, as well as by negative correlations between responsive channels and the negative bias that they introduce into all other channels.

To calculate the $R^2$, CCEPs were first averaged across trials at each channel. For each pair of channels, the mean time series at one channel, on the response interval between 10 and 300 ms, was used to predict the time series at the other channel using a linear model:

$$X_i(t) = \beta_1 + \beta_2 X_j(t),$$

Where $X_i(t)$ and $X_j(t)$ are the mean CCEP time series at the $i^{th}$ and $j^{th}$ channels, respectively. The $R^2$ was calculated and averaged over all pairs of $i$ and $j$, $i \neq j$, to produce a single value for each stimulation site. We calculated the average $R^2$ for all stimulation sites in all subjects, for five separate referencing conditions: without re-referencing (monopolar hardware-referenced), standard CAR using all channels, a naive low-variance CAR using the bottom 25% channels by mean cross-trial covariance, a second naive low-variance CAR using the bottom 50% channels by mean cross-trial covariance, and CARLA. To ensure that the results were not trivially driven by line noise, we applied notch filters to the re-referenced data, at 60, 120, and 180 Hz, prior to calculating $R^2$. We conducted independent Wilcoxon signed-rank tests to detect significant differences in mean $R^2$ pairwise between each of the initial four conditions and CARLA. The familywise error rate was controlled with Bonferroni correction across the four tests.

## 2.7. Electrode localization and visualization

Subject preoperative T1 MRIs were transformed into AC-PC space through affine transformations and trilinear voxel interpolation, such that the mid-sagittal plane lay on the y-z plane with anterior and posterior commissures on the y-axis (Huang et al., 2021). sEEG electrodes were localized from the postoperative CT scans and coregistered to the T1 MRIs (Hermes et al., 2010). This allowed for visualization of recording electrodes and stimulation sites on individual subject T1 MRI slices, in AC-PC space.

    Stimulation of different tissue types may result in different optima detected by CARLA. To test this possibility, the subject T1 MRIs were segmented using the autosegmentation algorithm in Freesurfer 7 (Dale et al., 1999). Gyral and sulcal labels were generated for each subject's pial surface, during autosegmentation, by aligning surface topology to the Destrieux cortical atlas. Each stimulation site was assigned the cortical or subcortical label corresponding to the most frequent voxel label within a 3 mm radius of the linearly interpolated center position between the stimulated electrode pair. To understand whether stimulation sites in different tissue types result in different thresholds on the number of common average channels, we used a one-way ANOVA to compare the mean fraction of channels included in the common average across cortical gray matter, subcortical gray matter, and white matter.

## 2.8. Data and code availability

The data that support the findings of this study are available in Brain Imaging Data Structure (BIDS) format, upon manuscript publication, on OpenNeuro. The code used to generate all results and figures is available on GitHub: https://github.com/hharveygit/CARLA_JNM. All simulated results can be reproduced with code alone prior to data release.

# 3. Results

## 3.1. Performance of CARLA on simulated CCEPs

We simulated sets of CCEP data with different percentages of channels containing true responses and tested the ability of CARLA to exclude responsive channels from the common average (Figure 4). Each simulated set contained 50 channels and 12 stimulation trials, with between 0 and 45 channels containing true responses. 30 unique sets were simulated at each level of responsiveness. We identified the optimal common average using two methods. The first method, labeled "Global Optimum", was to simply choose the *n* at the global maximum (least negative) $\zeta$. The second method, labeled "First-Peak Optimum" was to choose the *n* at the first local maximum in $\zeta$ before a significant decrease. A significant decrease was defined algorithmically as follows. The bootstrapped values for $\zeta$ at the first local maximum were pairwise subtracted from the bootstrapped values for $\zeta$ at the trough in *n* before the next greater $\zeta$ (See right-most example in Figure 4A) to yield bootstrapped differences. A left-tailed 95% confidence interval in difference was estimated, and the decrease was considered statistically significant if this confidence interval did not overlap 0. Local maxima are evaluated in order of increasing *n* until the first significant one is detected. The global maximum is used by default if $\zeta$ increases monotonically or if no significant decreases are encountered at any local maximum. We set a floor for optimal *n* at 10% of all channels with this approach, to avoid unstable fluctuations in local maxima when *n* is low. Alternatively, it may be valid to configure the floor to be an absolute number of channels (e.g., 10 channels).

Figure 4A shows example plots of $\zeta$ corresponding to single simulated datasets with 0%, 20%, 40%, 60%, and 80% of all channels containing responses. Figure 4B shows the simulated dataset matching Figure 4A, with responsive channels labeled in red. These examples match the median performance, across 30 sets, at each level of responsiveness. The optimal subset of channels, $U_n$, to use for CAR are labeled by arrowheads in the colors matching the two methods described above. For each of the two methods, we quantified the error, where a FN indicates a responsive channel erroneously included in the common average and a FP indicates a non-responsive channel erroneously excluded from the common average. Minimizing FN is relatively more important than minimizing FP, because the primary objective is to avoid introducing responsive channels into the common average. When the percentage of total channels responsive is low (i.e., below 60%), the two methods generally produced the same result, producing optimal *n* at the global maximum before the inclusion of responsive channels. With higher percentages of responsive channels, it became increasingly more likely for an early local maximum to be surpassed by the addition of more responsive channels. The addition of many more responsive channels, which are uncorrelated to each other on average, compensates for and divides the anticorrelations inflicted by any individual responsive channel. At 80% responsiveness, the global optimum produces a median common average with nearly all channels included. On the other hand, the first-peak optimum correctly stops at the 20% of channels with lowest covariance, right before the first responsive channel. FNs and FPs using each method across 30 simulated sets at each level of responsiveness are summarized with boxplots in Figure 4C. The median FN remained 0 when up to 84% of total channels were responsive using the first-peak optimum, but only up to 68% using the global optimum. FP remained low throughout all levels of responsiveness for both methods, with a maximum median value of 2.5 (out of 50 channels). Sensitivity and specificity are plotted in a similar format in Figure 4D. Sensitivity is the fraction of responsive channels excluded from the common average out of all channels excluded, while specificity is the fraction of non-responsive channels included in the common average out of all channels included. The mean sensitivity remains near 1 (perfect) up to 80% responsiveness for the first-peak optimum method. We focus on the first-peak optima throughout the rest of the analyses due to its greater sensitivity.

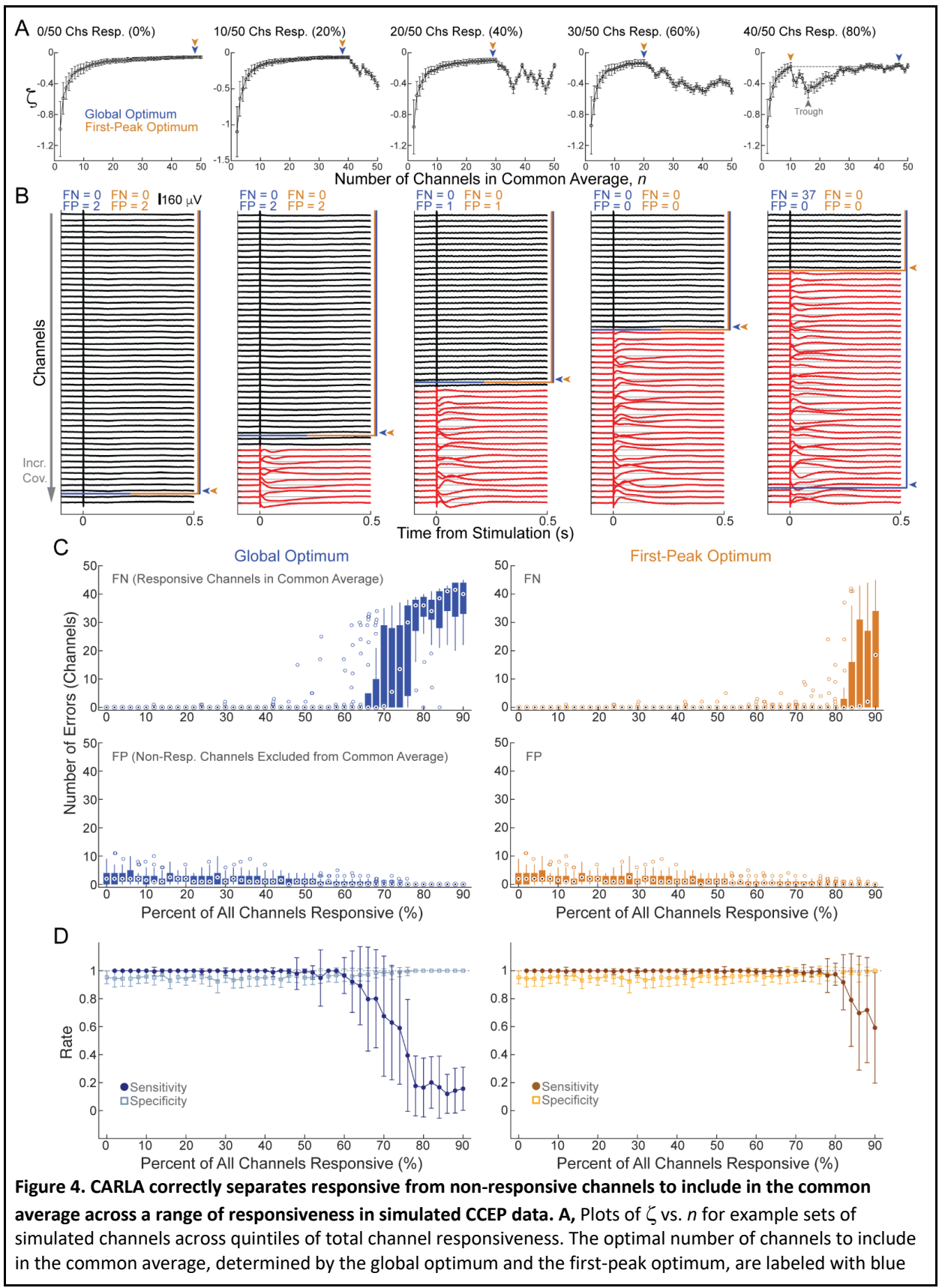

**Figure 4. CARLA correctly separates responsive from non-responsive channels to include in the common average across a range of responsiveness in simulated CCEP data. A,** Plots of $\zeta$ vs. $n$ for example sets of simulated channels across quintiles of total channel responsiveness. The optimal number of channels to include in the common average, determined by the global optimum and the first-peak optimum, are labeled with blue

and orange arrowheads, respectively. **B,** The mean signal across trials for each channel, corresponding to plots in A, sorted top-down by increasing mean cross-trial covariance. Signals in black contain no true response whereas signals in red contain a true response. The optimal $U_n$ used for the construction of the common average is labeled by the colored brackets (and the matching arrowhead at $n$), where blue and orange correspond to the global optimum and first-peak optimum methods, respectively. **C,** Boxplots showing the FN (top) and FP (bottom) for each method, across 30 simulated sets at each level of responsiveness. **D,** Mean sensitivity and specificity across the 30 simulated sets at each level of responsiveness. Error bars show standard deviation across sets.

## 3.2. CARLA results on real CCEP data

We next applied CARLA to real CCEP data where true responsiveness is unknown. Results from one stimulation site in subject 1 are presented in Figure 5. The global optimum and first-peak optimum occur at the same $n$ = 154 of 206 total channels (Figure 5B). The optimal $n$ coincides with a shoulder in the ordered covariances (Figure 5C), and the channels included in the common average appear visually less responsive than those that are not. ζ decreases precipitously at the end, which reflects the bias that would be introduced into the re-referenced signals from the last ~20 channels with the most extreme responses. Although notch filters alone can remove a significant amount of periodic line noise that is shared across channels, the signal quality is improved substantially by re-referencing, especially for individual trials (Figure 5F). No obvious bias is introduced into non-responsive channels by CARLA (E.g., channel "i" in Figure 5F). In contrast, a slight negative deflection is visibly introduced across channels when the same data is re-referenced by standard CAR instead (Figure S1).

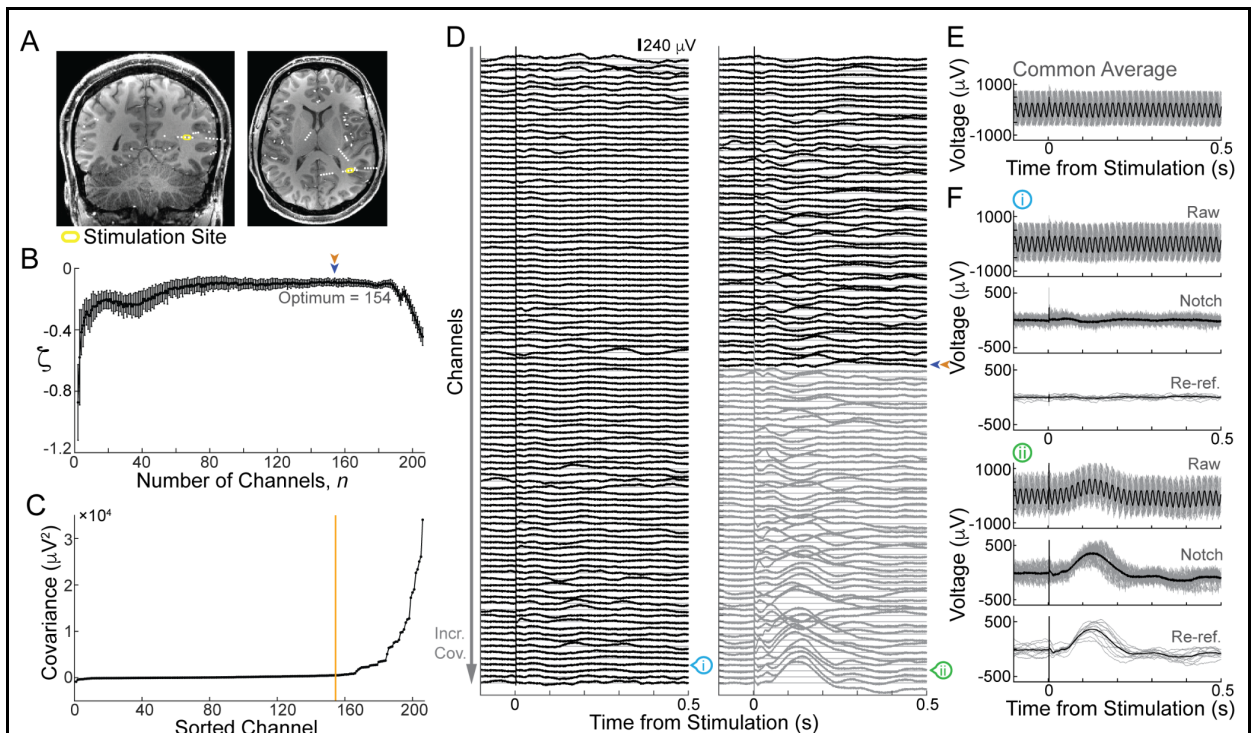

**Figure 5. CARLA constructs optimal common average from visually non-responsive channels in real CCEP data. A,** Coronal and sagittal T1-weighted MRI slices for subject 1 with electrodes within 4 mm of each slice overlaid.

The stimulation site, located in the white matter, is labeled in yellow. **B**, Plot of $\zeta$ vs. *n*. The global optimum (blue arrowhead) equals the first-peak optimum (orange arrowhead) and occurs at 154 out of 206 total recording channels. **C**, Mean cross-trial covariance for all channels sorted in increasing order, matching B. The orange vertical line is at the optimum *n*. **D**, Mean signal across the 12 trials for all channels, before re-referencing, plotted in order of increasing cross-trial covariance. Line noise was removed by notch filters at 60, 120, and 180 Hz to improve visibility. Black denotes the optimal subset of channels to be included in the common average and gray denotes channels to be excluded from the common average. **E**, The common average created from the optimal subset (no notch filter applied). Gray lines are individual trials, and the black line is the mean across trials. **F,** The channels labeled "i" (light blue) and "ii" (green) in **D**, before any processing (Raw), with line noise removed (Notch), and after re-referencing by CARLA, without line noise removal (Re-ref.).

Figure S2 summarizes the results from CARLA applied to CCEP data from a different stimulation site in subject 1, in the inferior temporal sulcus, as well as from stimulation sites in three other subjects. Similar to the performance on simulated data, CARLA identified the cutoff across a large range of stimulation responsiveness, from an optimal *n* of 56/108 channels (52%) in subject 4 to 195/205 channels (95%) in subject 1. Notably, CARLA identified a higher optimal *n* for the inferior temporal sulcus stimulation site than the white matter stimulation site in subject 1. This coincides with the visually apparent difference in global CCEP excitability between these two stimulation sites, with the white matter stimulation site producing a greater number of high-amplitude evoked potentials across measurement electrodes.

### 3.3. CARLA in the presence of a global signal

The common noise term discussed so far is present across all channels but unique in time to each trial. An assumption has been that there is no globally correlated structure across all channels and trials, which is generally the case. This assumption is challenged, however, when the reference electrode shows a modest response to stimulation. This can arise when the reference electrode is close enough to tissue affected by stimulation, which may be difficult to avoid entirely when an experiment includes many stimulation sites across the brain. For example, reference electrodes in the white matter that are unresponsive to stimulation in the cortex could show weak responses during thalamic stimulation due to the thalamus's widespread outputs. This occurred at a thalamic stimulation site in subject 1 and described in the results below.

Therefore, we introduce an additional conceptual component of the CCEP, termed the global signal, which is stimulation-locked and consistent across all channels and trials, when present (Figure S3). The global signal is visible in the mean signal across trials of most or all channels (Figure S3B). Its presence limits the functionality of a common average that simply excludes responsive channels with a statistically significant response (e.g., compared to a baseline interval), as these methods would spuriously exclude most or all channels from the common average.

CARLA makes no assumptions about the overall responsiveness of all channels and is well-suited to re-referencing these data. We present below the effect of CARLA on simulated and real CCEP data that contain a global signal component (Figure 6). In the simulated data, the global signal was modeled in the same way as the evoked potential (see Methods, above), but with an amplitude, $A$, that was four times lower (sampled uniformly at random between 20 and 30). This amplitude range was chosen to approximate real CCEP data, in which a global signal, when present, would be generally of much lower amplitude than the evoked potentials of interest when the reference electrode has been selected reasonably well.

Figure 6A presents an example data set, containing a global signal, of 50 simulated channels with 12 stimulation trials. 25 channels were truly responsive in addition to the global signal (red). On this simulated data set, CARLA identified an optimal subset of 24 non-responsive channels to use for re-referencing, for both optima (Figure 6A). Due to the global signal, responsive channels might rank earlier (lower mean cross-trial covariance) than non-responsive channels, and the frequency of this occurrence depends partially on the relative amplitude of the global signal. Overall, across 30 simulated sets at each level of responsiveness from 0 (0%) to 45 (90%) responsive channels, the first-peak optimum performance of CARLA remained stable and comparable to simulated data without a global signal. The median FN remained 0 up to 86% responsive channels, highlighting the robust sensitivity despite a global signal. The FP was marginally higher at all levels of responsiveness than data without a global signal, with a maximum median FP equal to 6, when 1 channel was responsive.

CARLA was applied to real CCEP data that visibly contained a global signal, resulting from a thalamic stimulation site in subject 1 and a hardware reference electrode in the white matter in a highly responsive hemisphere (Figures 6C-F). In this case, an identical response in the first 50 ms can be seen across almost all channels, but only a small subset of channels shows truly unique responses. On this data set, CARLA identifies an optimal subset of 181 out of 206 recording channels as nonresponsive, appropriately including most channels with the global signal in its construction of the common average. Non-responsive channels have the common signal effectively attenuated by re-referencing (Figure 6F, channel labeled "i"). Interestingly, a few channels do not contain the global signal before re-referencing. Upon closer inspection, these electrodes were all located in close physical proximity to the hardware reference electrode. It was likely that their recorded responses were similar to that of the reference electrode, and were therefore attenuated by the hardware reference. For instance, the channel marked "ii" (Figures 6F) was immediately adjacent to the reference electrode on the same sEEG lead, located 3.50 mm away. Re-referencing re-introduces the negative global signal as a true response in such channels. Overall, the signal quality is improved substantially by CARLA (Figure 6F), as was the case in the data without a global signal.

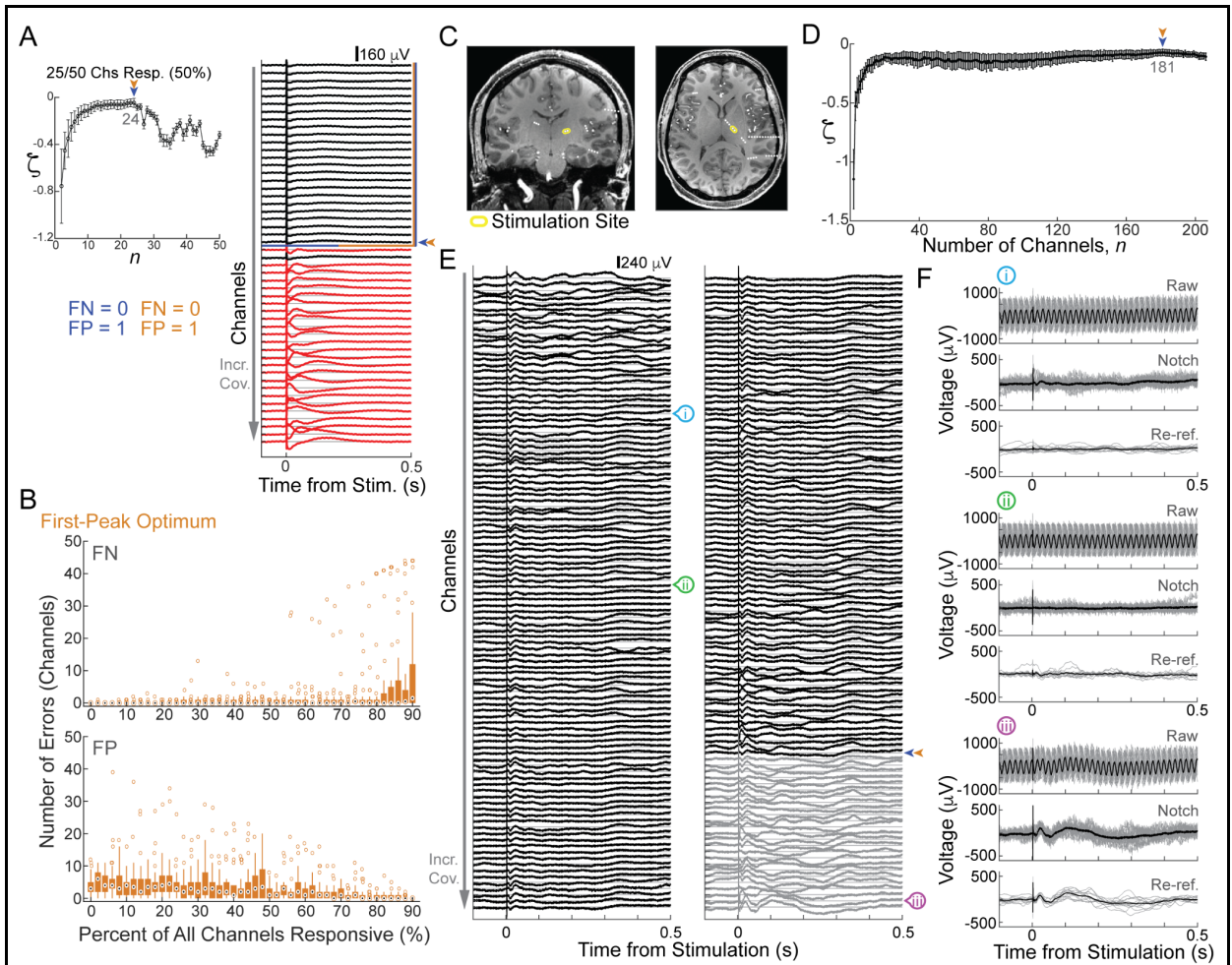

**Figure 6. CARLA identifies non-responsive channels to include in the common average in simulated and real CCEP data, in the presence of a weaker global signal. A,** Plot of $\zeta$ vs. $n$ (left) and the mean signal across trials for each channel, sorted top-down in increasing trial covariance (right), for a simulated set of CCEPs where 25 of 50 total channels are responsive. All channels, including non-responsive ones, contain a low-amplitude global signal that is consistent across trials. Channels in black contain no true response while channels in red contain a true response. The optimal number of channels to include in the common average, determined by the global optimum and the first-peak optimum, are labeled with blue and orange arrowheads, respectively, and are identical in this case. FN = False Negative, FP = False Positive. **B,** Boxplots showing the FN (top) and FP (bottom) for the first-peak optimum, across 30 simulated sets at each level of responsiveness. A unique random global signal is present for each simulated iteration. **C**, Coronal and sagittal T1-weighted MRI slices for subject 1 with electrodes within 4 mm of each slice overlaid. The stimulation site, located in the thalamus, is labeled in yellow. **D,** Plot of $\zeta$ vs. $n$. The global optimum (blue arrowhead) equals the first-peak optimum (orange arrowhead) and occurs at 181 out of 206 total recording channels. **E,** Mean signal across trials for all channels, before re-referencing, plotted in order of increasing cross-trial covariance. Line noise was removed by notch filters at 60, 120, and 180 Hz to improve visibility. Black denotes the optimal subset of channels to be included in the common average and gray denotes channels to be excluded from the common average. A global signal can be seen in nearly all channels. **F,** The channels labeled "i" (blue), "ii" (green), and "iii" (purple) in **E**, before any processing (Raw), with line noise removed (Notch), and after re-referencing by CARLA, without line noise removal (Re-ref.).

## 3.4. Size of the optimal common average across all stimulation sites

We calculated the optimal *n* identified by CARLA on all stimulation sites in all subjects (Figure 7). Some of these stimulation sites produced a visible global signal across all measurement channels while others did not. In subject 1, we observed that the cortical stimulation site (Figure S2) resulted in fewer responsive channels than the white matter stimulation site (Figure 5). Thus, we hypothesized that the degree of responsiveness might depend on the type of tissue in which the stimulation sites were located, and so we grouped the stimulation sites into three major tissue types: cortical gray, subcortical gray, and white matter. Cortical gray included all stimulation sites located in the gyri or sulci of the neocortex. Subcortical gray included all stimulation sites located in the thalamus, amygdala, and hippocampus. White matter included stimulation sites in the white matter only. The size of the optimal common average, expressed as a percent of all measurement channels analyzed for each stimulation site, ranged widely from 11% to 100%. The algorithmic floor of 10% imposed by the first-peak optimum was never incurred. The mean percentage did not differ significantly between tissue types within any individual subject (ANOVA, *p* > 0.05). The mean percentages across all stimulation sites in subjects 1 through 4 were 75%, 80%, 62%, and 65%, respectively.

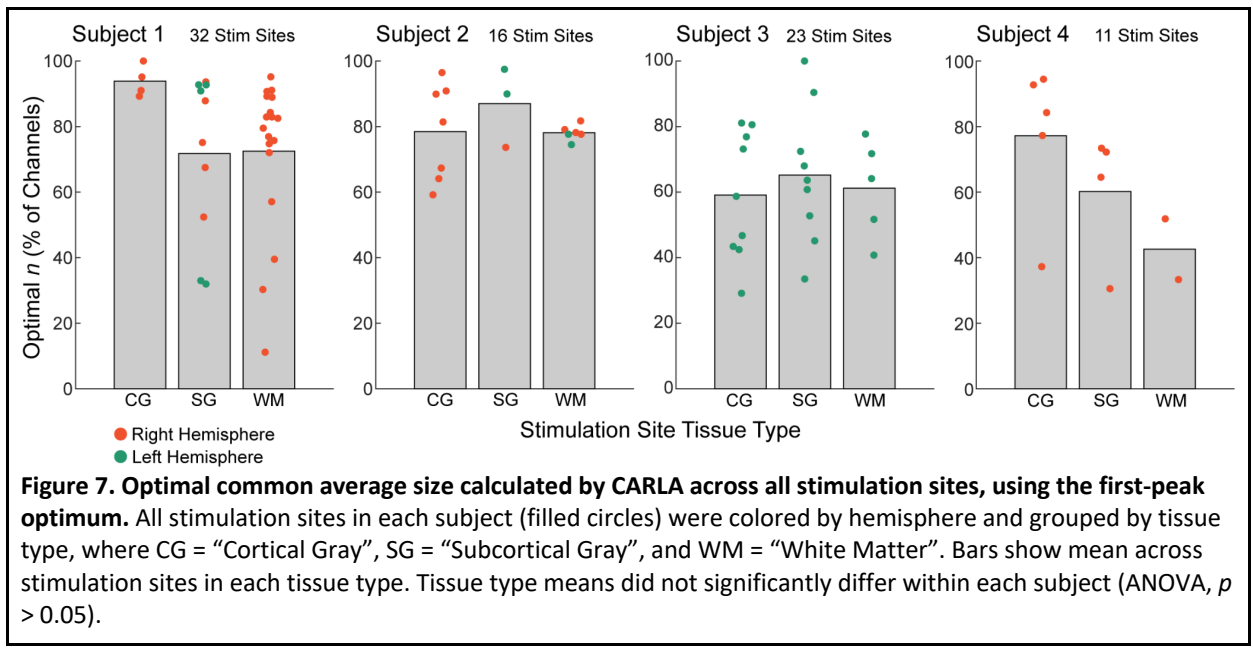

**Figure 7. Optimal common average size calculated by CARLA across all stimulation sites, using the first-peak optimum.** All stimulation sites in each subject (filled circles) were colored by hemisphere and grouped by tissue type, where CG = "Cortical Gray", SG = "Subcortical Gray", and WM = "White Matter". Bars show mean across stimulation sites in each tissue type. Tissue type means did not significantly differ within each subject (ANOVA, *p* > 0.05).

We compared the signal quality of CCEPs re-referenced by CARLA to four other referencing conditions, as quantified by the mean $R^2$ between all pairs of channels (Figure 8). Data were notch-filtered at 60, 120, and 180 Hz after re-referencing but before calculating $R^2$ to mitigate the contribution of periodic line noise. In well-referenced data, the mean cross-channel $R^2$ is expected to be low as there are few interdependencies between non-responsive channels. Figure 8A shows example $R^2$ matrices from stimulation site 1 in subject 1, which show the $R^2$ between all pairs of channels for each of the five referencing conditions. Without re-referencing, most channels are highly predictive of each other due to the large amount of common noise. CCEPs re-referenced by a standard CAR, where all measurement channels are used to construct the common average, show pairwise $R^2$ values that are substantially decreased over no re-referencing, but these $R^2$ values are on average still visibly greater than those from CCEPs re-referenced by the adjusted CARs. This is attributable to common bias introduced by responsive channels in the standard CAR (e.g., Figure S1). Figure 8B summarizes the mean cross-channel $R^2$, averaged over each matrix, for the 82 stimulation sites in all subjects for the five referencing conditions. CARLA produced significantly lower mean cross-channel $R^2$ than the no re-referencing, standard CAR, and bottom 25% CAR conditions (pairwise Wilcoxon signed-rank tests, $p < 0.001$, Bonferroni-corrected).

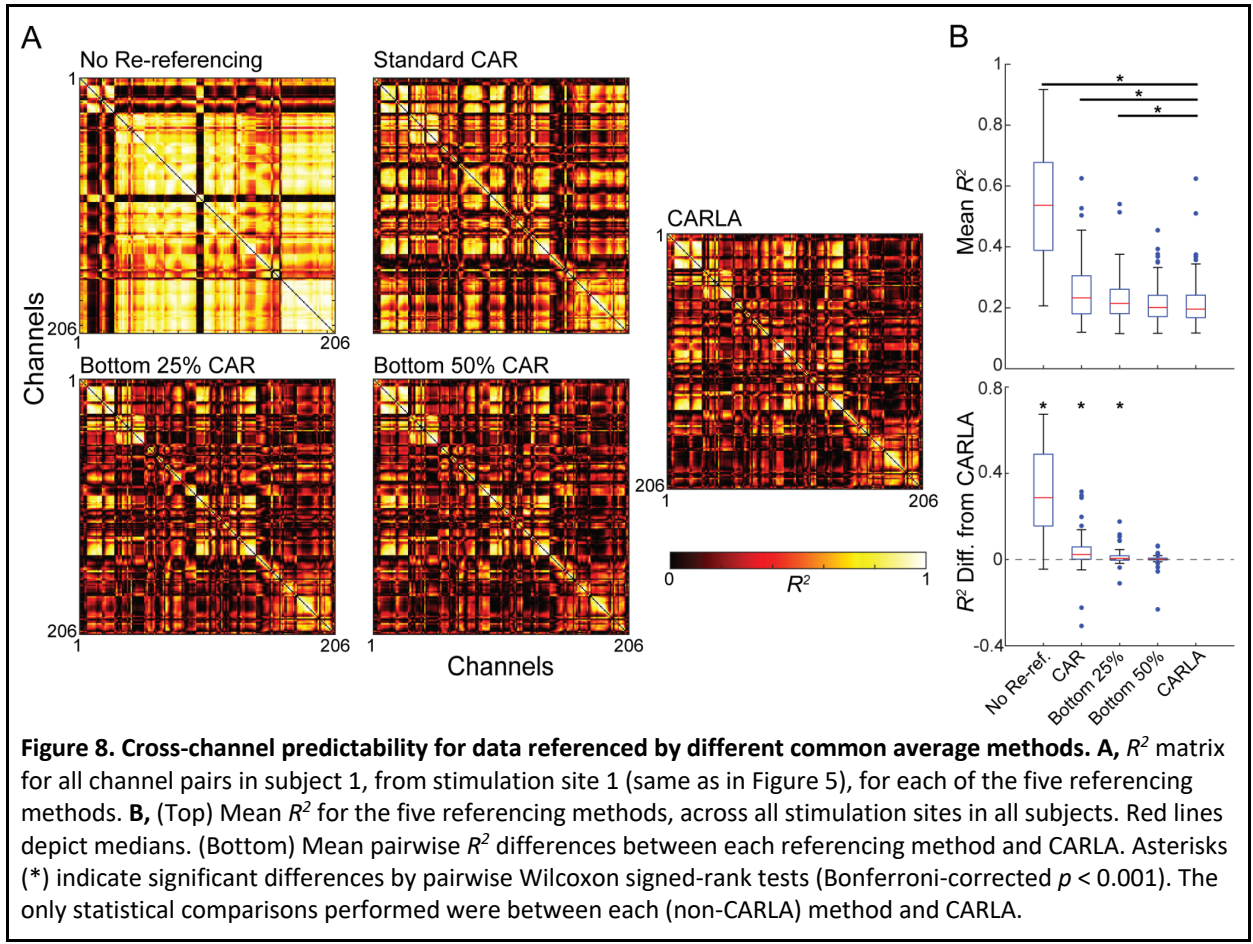

**Figure 8. Cross-channel predictability for data referenced by different common average methods. A,** $R^2$ matrix for all channel pairs in subject 1, from stimulation site 1 (same as in Figure 5), for each of the five referencing methods. **B,** (Top) Mean $R^2$ for the five referencing methods, across all stimulation sites in all subjects. Red lines depict medians. (Bottom) Mean pairwise $R^2$ differences between each referencing method and CARLA. Asterisks (*) indicate significant differences by pairwise Wilcoxon signed-rank tests (Bonferroni-corrected $p < 0.001$). The only statistical comparisons performed were between each (non-CARLA) method and CARLA.

CARLA did not differ significantly from the bottom 50% CAR condition, which suggests that a naive CAR using the bottom 50% of measurement channels by cross-trial covariance might be comparably effective to CARLA. However, this group-level similarity in average $R^2$ may depend on the data used and may be suboptimal for individual stimulation sites. Indeed, many stimulation sites showed an optimal common average size less than 50% of measurement channels (Figure 7). On the other hand, all but one stimulation site resulted in an optimal common average size greater than 25% of measurement channels. Since the $\zeta$ plots rapidly approached plateau for most stimulation sites (e.g., see Figures 5, S2), a conservative naive threshold of 25% may be reasonable. This threshold would prevent the introduction of bias from responsive channels in the vast majority of cases, at the cost of modestly increased inter-channel dependency.

# 4. Discussion

CAR is an effective offline re-referencing method that can improve signal quality in iEEG data, while preserving signal features that are distributed across local neighbor electrodes. CAR is somewhat uncommon in CCEP studies, given the potential introduction of bias from responsive channels into all other channels. In this paper, we have addressed this major shortcoming, by introducing CARLA, an adjusted CAR. CARLA constructs a common average using a low-variance subset of measurement channels, and it adapts the size of this subset by minimizing anticorrelation between any single channel and all other re-referenced channels. CARLA reduced noise without introducing bias in both simulated and recorded CCEP signals.

There are many ways to rank channels for inclusion in an adjusted CAR method. We chose to use the mean pairwise covariance between trials (when $K > 1$), rather than the simpler mean variance across trials. This feature has several advantages. Most importantly, the covariance between trials quantifies both average signal strength as well as reliability, which helps to prevent reliable responses from being included in the common average, not only high-amplitude ones. Furthermore, there are ($n^2 - n$)/2 values of pairwise covariance to average for $n$ trials, compared to $n$ values of variance, which results in a more reliable mean estimator for ranking. Finally, the covariance deviates minimally from the established metric of variance, as average covariance equals average variance in the case that all trials are identical.

CARLA employs a rather complex set of steps in developing the optimization statistic, $\zeta$. These steps were designed to maximize the robustness of the overall method. For each subset of $n$ channels $U_n$ to be considered as a common average, $\bar{z}_{i,n}$ is calculated as the average z-transformed Pearson's correlation between the $i^{th}$ channel in $U_n$ and all other channels in $U_n$ after re-referencing. Only correlations between channels within $U_n$ are calculated to avoid spurious positive or negative correlations with responsive channels that have not yet been considered. In other words, calculating correlations within $U_n$ ensures that channels are only weakly anticorrelated with each other so long as $U_n$ has not expanded to contain responsive channels. CARLA finds the minimum $\bar{z}_{i,n}$ from all channels in $U_n$ to assign as $\zeta$, rather than taking $\bar{z}$ calculated from the newest channel added to $U_n$, to maintain stability in $\zeta$ as $n$ increases. The minimum $\bar{z}_{i,n}$ is robust to spurious positive correlations between a newly-added responsive channel that is anticorrelated with a previously responsive channel (and thus positively correlated to the re-referenced signals, on average), which occurs when sEEG electrodes cross signal dipoles in the laminar cortex (Buzsáki et al., 2012; Huang et al., 2023; Mitzdorf, 1985). An example of this can be seen in Figure 5D, where many responsive channels are visibly anticorrelated with each other. Indeed, this approach permits CARLA to be agnostic to response shape. Although we tested CARLA on sEEG data, CARLA would in principle function robustly in the ECoG setting as well, where cortical evoked potential shapes are expected to be more well-preserved across measurement channels on the cortical surface. The output of CARLA is a simple percentile threshold on channels ranked by (co)variance per stimulation site. This output simplicity permits a single threshold to be chosen across all stimulation sites if desired *post-hoc*. Indeed, we provided justification for using fixed percentile thresholds of 25% or 50%. Finally, CARLA does not require extensive user fine-tuning of parameters, as the only adjustable parameters are the response interval and the significance threshold to detect a first-peak optimum.

It is not the goal of CARLA to completely minimize inter-channel dependency. Mean cross-channel $R^2$ was used in this analysis as a signal quality metric to compare CCEPs re-referenced by CARLA to those without re-referencing or re-referenced by other forms of CAR. Although CARLA yielded the lowest mean $R^2$ out of these methods, it would be inadequate to determine the optimal common average as simply as that which minimizes mean $R^2$. This is because responsive channels are often highly correlated with each other in CCEP data, and the goal of re-referencing is not to completely eliminate all correlative structure between channels. Local re-referencing methods would likely yield even lower mean $R^2$ (Li et al., 2018), but this is not ideal precisely because they eliminate signal features common to electrode neighbors.

CARLA presents several advantages over other possible re-referencing strategies on CCEP data. First, CARLA does not require a complete absence of stimulation-locked activity in channels to be included in the common average. This allows CARLA to function on CCEP data containing some amount of unwanted global signal or stimulation artifact. Moreover, defining which channels are responsive is a statistical problem that requires additional user-defined significance thresholds and possibly long periods of baseline data to compare against. Alternatively, one might attempt to determine the optimal threshold by finding a shoulder in the ranked channel covariances alone, such as that in Figure 5C. This would not be trivial to generalize to all data sets, as the ranked covariances can take on different (e.g., sigmoidal) shapes and increase with varying slope. Some referencing methods employ dimensionality reduction techniques such as principal and independent component analysis (Alexander et al., 2019; Michelmann et al., 2018; Uher et al., 2020). These methods may produce unstable solutions that differ significantly for similar input data and require careful inspection of desired components. Average-based references like CARLA tend to exhibit greater stability and automation. CARLA is also indifferent to baseline offsets in the data, as both the covariance used to rank channels and the Pearson correlation used for optimization are calculated on mean-centered values. Finally, in contrast to white matter averaging methods, CARLA is entirely iEEG data-driven and does not require anatomical data.

This method is optimized for evoked potentials in single pulse stimulation data, and it may not be optimal for other desired signal features, such as oscillations and broadband power changes. However, it is advantageous in that data re-referenced by CARLA can then be easily re-referenced again by local referencing methods (e.g., bipolar) with no loss of information; the subtracted common average cancels out between neighboring electrodes. This permits multiple signal features to be efficiently extracted in series on a linear pipeline. One assumption in both CARLA and local referencing methods is that all measurement electrodes have identical impedance, as otherwise the common noise may be differentially amplified across electrodes. In our data this assumption is generally well met, and we exclude channels with visibly different amplitudes before preprocessing. If this assumption is not met, the data might be better referenced by a scalable noise term using a regression-based approach. Noise may be common to subgroups of measurement electrodes, such as those connected to a single connector box (e.g., Natus breakout box or G.Tec G.HEADbox). In these circumstances, CARLA might be more appropriately performed at the subgroup level rather than across all measurement electrodes at once. The overall accuracy of CARLA may further depend on the overall signal-to-noise ratio of CCEPs, as well as the variability of evoked potential amplitudes across responsive channels. Anticorrelations introduced by CCEPs with lower signal-to-noise ratio would be more subtle and impair sensitivity. Similarly, if a global signal with comparable amplitude to evoked potential responses is present, the relative ranking of channels by covariance can be less reliable. This could result in responsive channels ranked earlier in the construction of $U_n$ and a smaller optimal $n$ on average (more FNs). This potential limitation is mitigated by careful selection of the reference electrode to be as electrophysiologically neutral as possible during data collection.

In conclusion, CARLA is an adjustable common average re-referencing method that improves the signal quality of evoked potentials in CCEP data. We have demonstrated that CARLA accurately excludes truly responsive channels in simulated CCEP data, improves signal quality in real CCEP data collected from four human subjects, and even functions in the presence of an unwanted global signal.

# Acknowledgements

We are grateful for the participation of the patients in this study, and for the assistance of Cindy Nelson, Karla Crockett, and other staff at Saint Mary Hospital, Mayo Clinic, Rochester, MN. Research reported in this publication was supported by the National Institute of Mental Health under Award Number R01MH122258, by the National Institute of General Medical Sciences under Award Number T32GM065841, and by the American Epilepsy Society under award number 937450. We thank Daniel A.N. Barbosa for helpful discussions. The content is solely the responsibility of the authors and does not represent the official views of the National Institutes of Health.

# Supplementary Figures

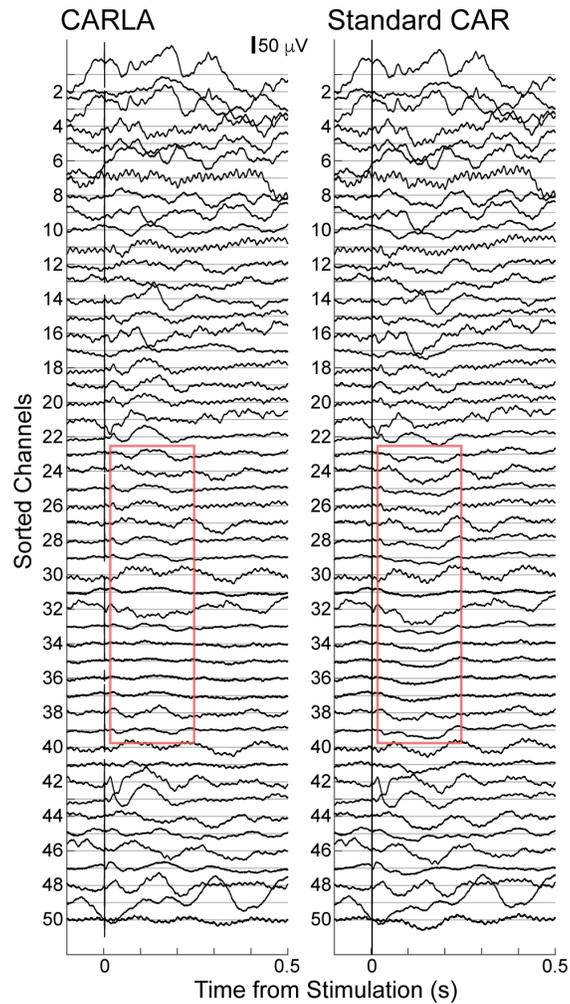

**Figure S1. Bias is introduced into channel signals by standard CAR but not by CARLA.** Mean signal across trials magnified for the first fifty sorted channels in Figure 5D, after re-referencing with CARLA (left) or standard CAR (right). A slight negative deflection between 0 and approximately 200 ms post-stimulation, corresponding to bias introduced by the common average, is visible across all channels in the standard CAR-referenced data but not the CARLA-referenced data. The red box highlights channels where this difference is most easily appreciable.

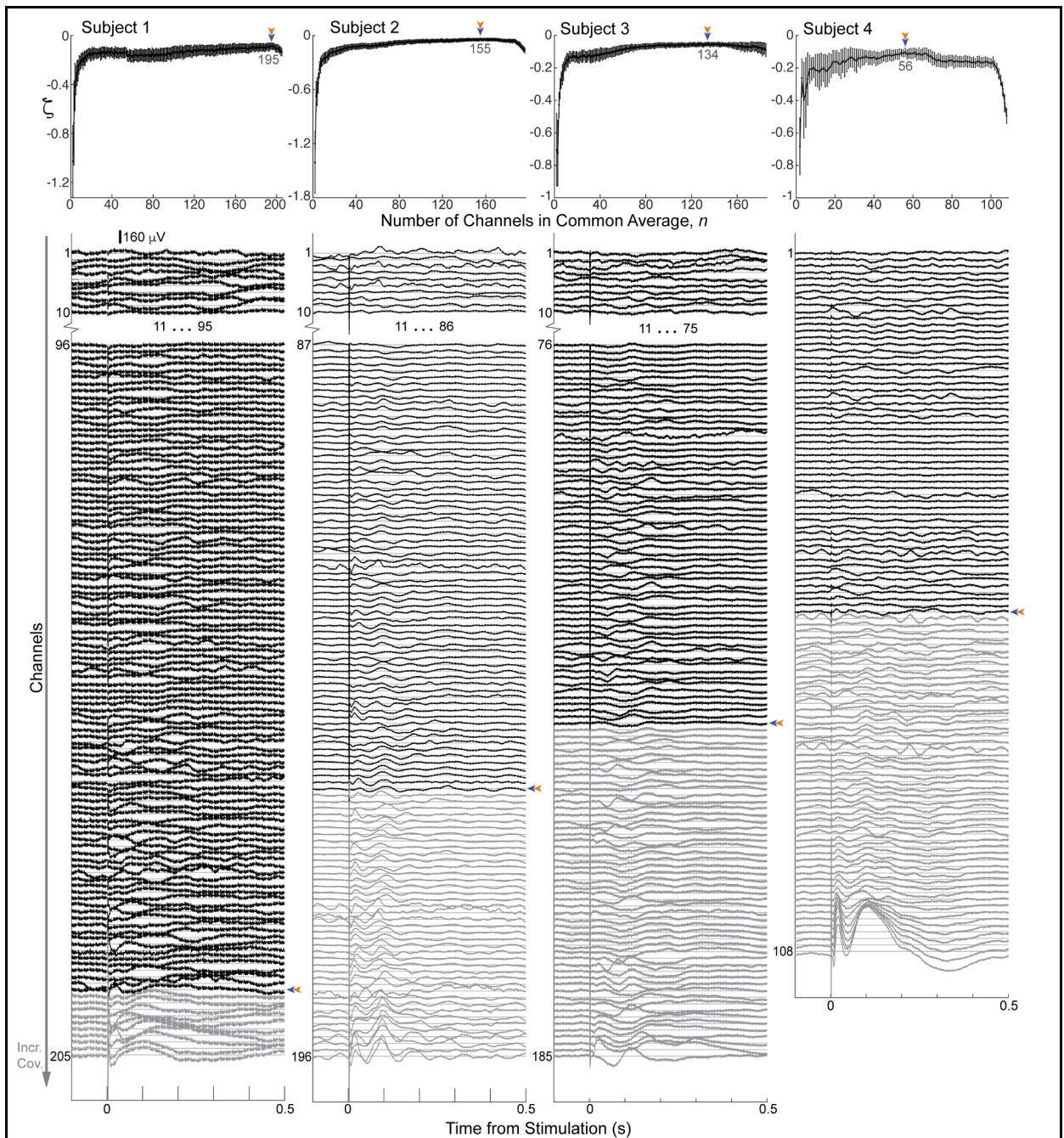

**Figure S2. Application of CARLA to CCEP data from stimulation sites in four subjects.** Each column corresponds to a stimulation site in a different subject. There were 12 trials performed in each case. Stimulation sites were located in the white matter for subjects 2 and 4, in the inferior temporal sulcus for subject 1 (different site than shown in Figure 5), and in the amygdala for subject 3. Top: $\zeta$ vs. *n* plot for each stimulation site, as in Figure 5. In each case, the global optimum (blue arrowhead) equals the first-peak optimum (orange arrowhead). Bottom: Mean signal across trials for all channels, before re-referencing, plotted in order of increasing cross-trial covariance. Line noise was removed by notch filters at 60, 120, and 180 Hz to improve visibility. Black denotes the optimal subset of channels to be included in the common average and gray denotes channels to be excluded from the common average. In subjects 1 through 3, a middle segment of the nonresponsive channels is omitted (indicated by the ellipses) to better visualize the channels immediately before and after the optimal cutoff.

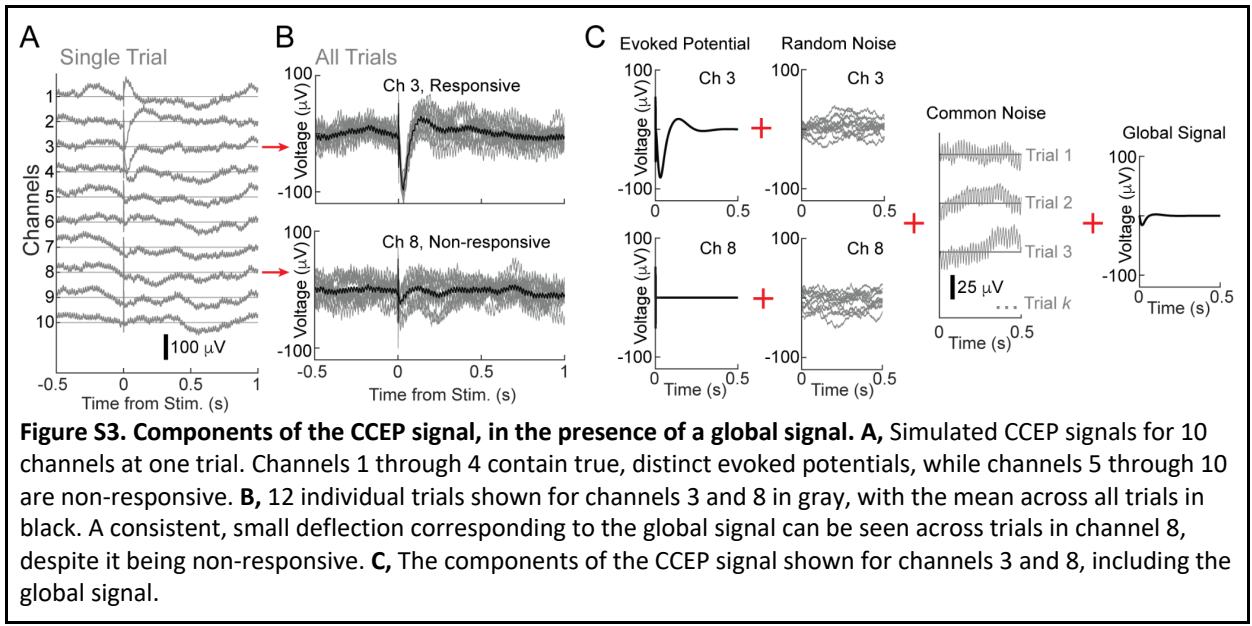

**Figure S3. Components of the CCEP signal, in the presence of a global signal. A,** Simulated CCEP signals for 10 channels at one trial. Channels 1 through 4 contain true, distinct evoked potentials, while channels 5 through 10 are non-responsive. **B,** 12 individual trials shown for channels 3 and 8 in gray, with the mean across all trials in black. A consistent, small deflection corresponding to the global signal can be seen across trials in channel 8, despite it being non-responsive. **C,** The components of the CCEP signal shown for channels 3 and 8, including the global signal.